\documentclass[12pt]{iopart}

\usepackage{iopams}
\usepackage{graphicx}

\begin{document}

\title[Unstable spectrum of a relativistic electron beam...]{Unstable spectrum of a relativistic electron beam interacting with a quantum collisional plasma: application to the Fast Ignition Scenario}

\author{A. Bret, F. J. Mar\'{\i}n Fern\'{a}ndez, J.M. Anfray}

\address{ETSI Industriales, Universidad de Castilla-La
Mancha, 13071 Ciudad Real, Spain}
\ead{antoineclaude.bret@uclm.es}

\maketitle

\begin{abstract}
Quantum and collisional effects on the unstable spectrum of a relativistic electron beam-plasma system are investigated through a two-fluids model. Application is made to the near target center interaction of the relativistic electron beam in the Fast Ignition Scenario. Partial degeneracy effects are found negligible while the most influential factors are the beam temperature and the electron-ion collision frequency of the plasma. The introduction of the latter triggers some oblique unstable modes of much larger wave length than the collisionless ones. Transition from the collisionless regime to the resistive one is thus documented and found discontinuous.
\end{abstract}

\pacs{52.57.Kk,52.35.Qz}

\maketitle

\section{Introduction}
The Fast Ignition Scenario (FIS) for Inertial Confinement Fusion assumes that a pre-compressed Deuterium-Tritium target is ignited by a Petawatt Laser generated relativistic electron beam, originated near the border of the pellet \cite{Tabak,Tabak2005}. This scenario offers a very interesting theoretical setting to plasma physics, as it includes the interaction of a relativistic electron beam with a plasma which density varies over several orders of magnitude. When the electron beam reaches the center of the pre-compressed pellet, the plasma density is so high that quantum effects are to be accounted for. Interestingly, if beam-plasma instabilities are not welcome at the beginning of the beam travel to the core, they may not be so deleterious at the end of it because they contribute to the beam stopping and energy deposition. Much efforts have been devoted so far to the beam-plasma interaction physics involved in the whole process \cite{Silva2002,Tatarakis,Honda2004,hill2005}, but the investigation of quantum effects is still in its infancy. As far as collisional effects are concerned, recent works focused on the filamentation instability (see Refs. \cite{hao2008,cottrill2008} and references therein). While this instability is expected to govern the system at the beginning of the beam path because of the high beam to plasma density ratio \cite{BretPRL2008}, the low density ratio near the target center would rather point towards an ``oblique modes'' driven system, where the dominant mode wave vector makes an oblique angle with the beam flow. Regarding these modes, quantum effects have never been assessed, collisional ones are almost unknown and the interplay between these two remains unidentified. Finally, it has been now established that beam temperature is a key parameter \cite{Silva2002,BretPRE2004,BretPRE2005} when describing the unstable spectrum.

Our intent is to construct a theory accounting for all the aforementioned effects in order to identity the most influential ones, and focus on them in later works.  The model presented here includes therefore the following features: 1) a relativistic beam with transverse and parallel temperatures, 2) a weakly degenerate and 3) collisional plasma, and finally, 4) the calculation of the whole unstable spectrum  in order to spot the most unstable modes in this setting. We  thus implement a quantum collisional two-fluids model, as previous works \cite{BretPoPFluide} demonstrated that such formalism is able to reproduce the results from a waterbag kinetic theory \cite{BretPRE2004,BretPRE2005}. Note that on the one hand, the fluid treatment of thermal effects requires sub-relativistic temperatures (see remark after Eq. \ref{eq:forceb}) while on the other hand, thermonuclear burn of the target should deeply modify the physics considered in the sequel. The process considered here is thus the beam interaction with the pre-ignited and pre-compressed FIS target core.

\section{Basic equations}
We consider the interaction of an homogenous and infinite electron beam of density $n_{b0}$ and velocity $v_{b0}$ with a plasma of density $n_{p0}$. A plasma return current at velocity $v_{p0}$ cancels exactly the beam current with $n_{b0}v_{b0}=n_{p0}v_{p0}$. Ions are considered a fixed neutralizing background. Furthermore, our focus on the near target center interaction implies $n_{b0}\ll n_{p0}$. Both beam and plasma electrons share the same conservation equation,
\begin{equation}\label{eq:conser}
   \frac{\partial n_j}{\partial t}+\nabla\cdot(n_j \mathbf{v}_j)=0,
\end{equation}
but we use quite different Euler equations for the two species.

On the one hand, we assume a relativistic beam which density is low enough for quantum effects to be neglected. We also neglect here all sort of collisions due to the relativistic velocity involved; collisionality is here restricted to the plasma species \cite{gremillet2002,cottrill2008}. The beam Euler equation reads therefore
\begin{equation}\label{eq:forceb}
   \frac{\partial \mathbf{p}_b}{\partial
   t}+(\mathbf{v}_b\cdot\nabla)\mathbf{p}_b=-q\left(\mathbf{E} + \frac{\mathbf{v}_b\times
   \mathbf{B}}{c}\right)
   -\frac{\nabla P_b}{n_b},
\end{equation}
where $q>0$ is the electron charge and $\mathbf{p}_b=\gamma m \mathbf{v}_b$ the relativistic momentum. The kinetic pressure term will be dealt with according to the guidelines set in Ref. \cite{BretPoPFluide}: The pressure gradient is first expressed as $\nabla P_b = 3 k_B T_b\nabla n_b$ \cite{Sentoku2000,Honda2004} before a temperature tensor is introduced in the linearized equation (see Eqs. \ref{eq:forcebL},\ref{eq:TensorT} and more comments bellow). Such an adiabatic treatment requires sub-relativistic temperatures \cite{Siambis79,Pego84}. While this is not a strong requirement for the plasma, such condition is more stringent for the beam.

On the other hand, plasma electrons are drifting with the non-relativistic velocity $v_{p0}=(n_{b0}/n_{p0})v_{b0}\ll c$. They suffer collisions with the background plasma and are weakly degenerate. Thus, we use for them the following modified Euler equation
\begin{eqnarray}\label{eq:forcep}
   \frac{\partial \mathbf{v}_p}{\partial
   t}+(\mathbf{v}_p\cdot\nabla)\mathbf{v}_p=&-&\frac{q}{m}\left(\mathbf{E} + \frac{\mathbf{v}_p\times\mathbf{B}}{c}\right)-\nu \mathbf{v}_p\nonumber\\
   &+&\frac{\hbar^2}{2m^2}\nabla\left(\frac{\nabla^2\sqrt{n_p}}{\sqrt{n_p}}\right)
   -\frac{n_{p0}v_{Tp}^2}{n_p}G\nabla\left(\frac{n_p}{n_{p0}}\right)^3.
\end{eqnarray}
The first two terms are just the Lorentz and the friction force where $\nu$ is the collision frequency between the plasma return current and the background ions. The third and fourth terms have been proposed recently as the quantum correction to the fluid equations for a finite temperature Fermi plasma  \cite{Eliasson2008}. They result from the combination of the so-called Bohm pressure term \cite{Mandredi2005,Haas2000,Haas2005} for a completely degenerate jellium, with a correction proportional to the thermal velocity $v_{Tp}^2=k_BT_p/m$ accounting for the finite plasma temperature $T_p$. Equation (\ref{eq:forcep}) can be seen as a combination of Eq. (19) in Ref. \cite{Eliasson2008} and Eq. (19) in Ref. \cite{Haas2005} where spin and relativistic effects are neglected. The function $G$ in the equation above reads \cite{Eliasson2008},
\begin{equation}\label{eq:G}
    G =\frac{   \mathrm{Li}_{5/2}(-e^{\beta\mu}) }{   \mathrm{Li}_{3/2}(-e^{\beta\mu})  },
\end{equation}
where $\mu$ is the chemical potential, $\beta^{-1}=k_BT_p$ and $\mathrm{Li}_s$ is the polylogarithm function,
\begin{equation}\label{eq:polylog}
    \mathrm{Li}_s(z)=\sum_{k=1}^\infty\frac{z^k}{k^s}.
\end{equation}
 Denoting $E_F$ the Fermi energy of the electron gas, $G \sim \frac{2}{5}\beta E_F$ for $k_BT_p\ll E_F$. For $E_F< 2 k_BT_p$,
 \begin{equation}\label{eq:Gapprox}
    G\sim 1+\frac{1}{3\sqrt{2\pi}}\left(\frac{E_F}{k_BT_p}\right)^{3/2}.
 \end{equation}
 A core density of $10^{26}$ cm$^{-3}$ yields $E_F=0.78$ keV. With a core temperature of a few hundreds of eV, we find that neither the classical $k_BT_p\gg E_F$ nor the cold jellium approximations $k_BT_p\ll E_F$ are really valid for the electron gas because its temperature is quite close to its Fermi energy. Partially degenerate thermal corrections to the Bohm pressure are thus required in Eq. (\ref{eq:forcep}).

The beam is relativistic but classical while the return current is non-relativistic but quantum. We can thus account for both relativistic and quantum effects through Eqs. (\ref{eq:forceb},\ref{eq:forcep}) and do not need quantum relativistic corrections in Eq. (\ref{eq:forcep}).

In order to compute the dispersion equation, Eqs. (\ref{eq:conser},\ref{eq:forceb},\ref{eq:forcep}) are linearized assuming small perturbations of the equilibrium variables proportional to $\exp(i\mathbf{k}\cdot \mathbf{r}-i\omega t)$ where $i^2=-1$. In the present relativistic diluted beam regime, we expect oblique modes to govern the system \cite{BretPRL2008}. We thus consider both  parallel and  perpendicular components of the wave vector to make sure the fastest growing modes are not overlooked. Choosing the $z$ axis as the beam direction, we can set $\mathbf{k}=(k_x,0,k_z)$ without loss of generality. The linearized equations are
\begin{equation}\label{eq:conserL}
   n_{j1}=\frac{\mathbf{k} \cdot \mathbf{v}_{j1}}{\omega-\mathbf{k}\cdot \mathbf{v}_{j0}},
\end{equation}
for the perturbed beam and plasma densities $n_{j1}$, and for the beam
\begin{eqnarray}\label{eq:forcebL}
im\gamma_b(\mathbf{k}\cdot \mathbf{v}_{b0}-\omega)\left[\mathbf{v}_{b1}+\frac{\gamma_b^2}{c^2}(\mathbf{v}_{b0}\cdot \mathbf{v}_{b1})\mathbf{v}_{b0}\right]=&-&q\left(\mathbf{E}_1+\frac{\mathbf{v}_{b0}\times \mathbf{B}_1}{c}\right)\nonumber\\
&-&3i\frac{n_{b1}}{n_{b0}}k_B\mathbf{T}_b\cdot \mathbf{k},
\end{eqnarray}
where $\mathbf{T}_b$ is the temperature tensor
\begin{equation}
\mathbf{T}_b=\left(
\begin{array}{ccc}
T_{b\perp} & 0 & 0 \\
0 & T_{b\perp} & 0 \\
0 & 0 & T_{b\parallel}
\end{array}
\right) .  \label{eq:TensorT}
\end{equation}
Although some covariant \cite{SilvaBAP,Schlickeiser2004,Schaefer2005,Achterberg2007} or kinetic \cite{BretPRL2008,BretPRL2005} treatment  of the beam temperature would be more appropriate because of the relativistic regime involved here, it was proved in Ref. \cite{BretPoPFluide} that the present formalism can reproduce the results from a waterbag kinetic theory, providing the perpendicular beam temperature parameter $T_{b\perp}$ is re-scaled by a factor $1/\sqrt{\gamma_b}$. At any rate, both waterbag kinetic theory and the present model are quite limited in their treatment of temperature. Nevertheless, waterbag theories are quite common \cite{yoon,Silva2002,BretPRE2004} as a first approach to kinetic effects, especially in the relativistic regime. The linearized Euler equation for the plasma reads,
\begin{eqnarray}\label{eq:forcepL}
i m (\mathbf{k}\cdot \mathbf{v}_{p0}-\omega)\mathbf{v}_{p1}=&-&q\left(\mathbf{E}_1+\frac{\mathbf{v}_{p0}\times \mathbf{B}_1}{c}\right)
-m\nu(\mathbf{v}_{p0}+\mathbf{v}_{p1})\\
&-&i\frac{\hbar^2 k^2}{4 m^2}\frac{n_{p1}}{n_{p0}}\mathbf{k}+3iGv_{Tp}^2\frac{n_{p1}}{n_{p0}}\mathbf{k}\nonumber
\end{eqnarray}
From this stage, Eqs. (\ref{eq:conserL},\ref{eq:forcebL},\ref{eq:forcepL}) are solved to express the perturbed velocities $\mathbf{v}_{p1}$ and $\mathbf{v}_{b1}$ in terms of $\mathbf{E}_1$ and $\mathbf{B}_1$. The first order magnetic field is then eliminated through $\mathbf{B}_1=(c/\omega)\mathbf{k}\times\mathbf{E}_1$, yielding an expression of the current,
\begin{equation}\label{eq:J}
\mathbf{J}=\sum_{j=b,p}n_{j0}\mathbf{v}_{j0}+n_{j0}\mathbf{v}_{j1}+n_{j1}\mathbf{v}_{j0},
\end{equation}
in terms of $\mathbf{E}_1$ only. Finally, the equation above is inserted into a combination of Maxwell-Faraday and Maxwell-Amp\`{e}re equations,
\begin{equation}\label{eq:Maxwell}
    \frac{c^2}{\omega^2}\mathbf{k}\times(\mathbf{k}\times \mathbf{E}_1)+\mathbf{E}_1+\frac{4 i \pi}{\omega}\mathbf{J}_1(\mathbf{E}_1)=0,
\end{equation}
which gives the dielectric tensor. This tensor has been here symbolically computed with the \emph{Mathematica} Notebook described in Ref. \cite{BretCPC} and is provided as ``Supplementary Material''to the paper. It is expressed in terms of the following dimensionless variables,
\begin{equation}\label{eq:param1}
    \alpha=\frac{n_{b0}}{n_{p0}},~~\mathbf{Z}=\frac{\mathbf{k}v_{b0}}{\omega_p},~~\beta=\frac{v_b}{c},
    ~~\tau=\frac{\nu}{\omega_p},~~\rho_{b\perp,\parallel}^2=\frac{3 k_BT_{b\perp,\parallel}}{m v_{b0}^2},
\end{equation}
where $\omega_p^2=4\pi n_{p0}q^2/m$ is the plasma electronic frequency, and
\begin{equation}\label{eq:param2}
    \Theta_c=\left(\frac{\hbar\omega_p}{2mc^2}\right)^2,~~\Theta_{T}=3G\left(\frac{v_{Tp}}{v_{b0}}\right)^2,
\end{equation}
where the parameter $G$ is given by Eq. (\ref{eq:G}). While $\Theta_c$ measures the strength of the zero temperature Bohm pressure term, $\Theta_{T}$ accounts for finite temperature effects. In this respect, the bridge between the completely degenerate and the classical cases is established through this parameter \cite{Eliasson2008}.

\begin{figure}[t]
\begin{center}

\includegraphics[width=\textwidth]{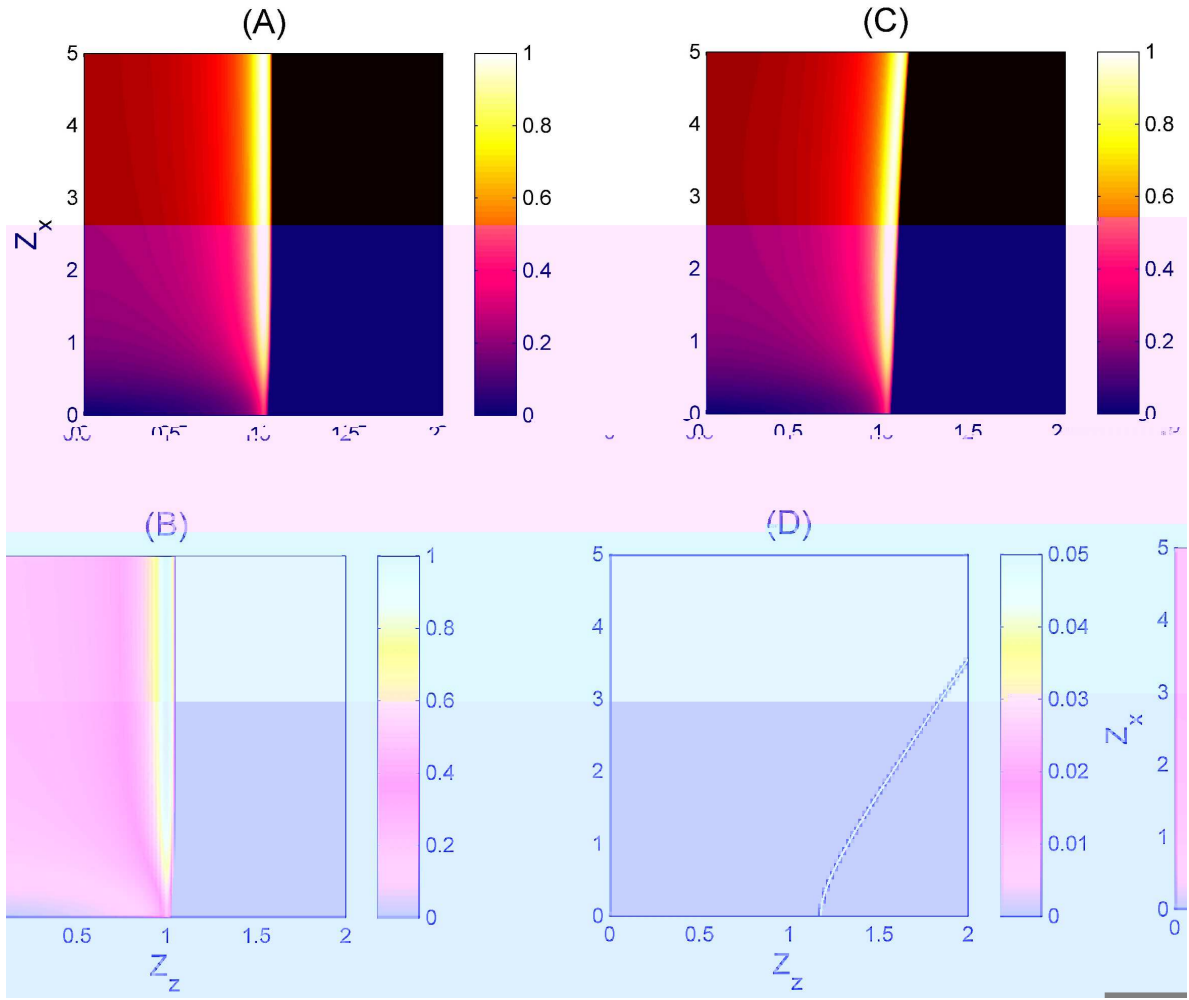}
\end{center}
\caption{(Color online) Growth rate normalized to the cold collisionless one given by Eq. (\ref{eq:GRcold},\ref{eq:GRColdNum}), in terms of $\mathbf{Z}=\mathbf{k}v_{b0}/\omega_p$. (a) Cold classical collisionless case. (b) Former case plus Bohm pressure term. (c) Former case plus quantum thermal corrections. (d) Thermal effects included, plus all quantum corrections. Collision frequency is zero ($\tau=0$) for all figures.} \label{fig:1}
\end{figure}

\section{\label{sec:Q}Quantum and thermal effects}
We start from the most simple cold-beam/cold-collisionless-plasma case for which the maximum growth rate $\delta_M$ is reached for an oblique wave $\mathbf{Z}_M$ vector with \cite{fainberg},
\begin{equation}\label{eq:GRcold}
    \frac{\delta_M}{\omega_p} = \frac{\sqrt{3}}{2^{4/3}}\left(\frac{\alpha}{\gamma_b}\right)^{1/3}~~Z_{zM}\sim 1,~~Z_{xM}\gg 1.
\end{equation}
In the limit of this cold regime, there is not one single most unstable mode but a continuum of oblique modes growing at the same rate \cite{califano3}. Because we deal with multiple effects at once, we focus on some typical beam and plasma parameters for FIS, and look at the influence of beam temperature, plasma degeneracy and collisionlality on the most unstable mode as well as its growth rate. We thus consider from now on parameters values extracted at peak compression from a global simulation performed by Ren \emph{et. al.} \cite{ren2006},
\begin{eqnarray}\label{eq:paramnum}
    n_b=10^{22}~\mathrm{cm}^{-3},~~\gamma_b=3,~~T_{b\perp,\parallel}=100~\mathrm{keV},\nonumber\\
    n_p=10^{26}~\mathrm{cm}^{-3},~~T_p=1~\mathrm{keV},~~\nu=0.4\omega_p.
\end{eqnarray}
Other simulations consider the same values for the core parameters \cite{Honrubia2006}. The fixed dimensionless parameters in Eqs. (\ref{eq:param1}) are thus,
\begin{equation}\label{eq:param2num}
    \alpha=10^{-4},~~\rho_{b\perp,\parallel}=0.76,~~
    \Theta_c=1.3\times 10^{-7},~~\Theta_{T}=6.5\times 10^{-3},
\end{equation}
while the maximum growth rate from Eq. (\ref{eq:GRcold}) is
\begin{equation}\label{eq:GRColdNum}
\delta_M=0.022\omega_p.
\end{equation}

\subsection{Quantum effects}
Starting from the cold, classical collisionless growth rate map on Fig. \ref{fig:1}(a), we successively add ``cold'' quantum effects on Fig. \ref{fig:1}(b), then partial degeneracy correction on Fig. \ref{fig:1}(c) and beam temperature on Fig. \ref{fig:1}(d). It is clear from Fig. \ref{fig:1}(b) that the Bohm pressure term hardly changes the global picture whereas Fig. \ref{fig:1}(c) shows more influence from the partial degeneracy correction, although the maximum growth rate is not affected.

Quantum thermal corrections arising from the ``$G$'' term in Eq. (\ref{eq:forcep}) can be compared with the so-called Bohm pressure term. Considering the unstable spectrum emphasizes wave vectors with $k\sim\omega_p/c$, and comparing the two quantum terms,  the Bohm term is found secondary while
\begin{equation}\label{eq:compaQ}
    \Theta_c \ll \Theta_T,
\end{equation}
In view of the numerical values given by Eqs. (\ref{eq:param2num}), the condition above is largely fulfilled. The plasma is  therefore hot enough for the Bohm pressure term to be negligible. The two quantum parameters would be comparable for $\Theta_T$ about $5\times 10^4$ smaller. Because $\Theta_T\propto v_{Tp}^2\propto T_p$, this would require $T_p\sim 1/50$ eV.

The fluid quantum theory of the filamentation instability \cite{BretPoPQuantum2007,BrePoPQB0} showed that quantum effects reduce small wave length instabilities through quantum interferences. We find here a similar trend all over the unstable spectrum. By lowering the growth rates at large $k_x$ (i.e., $k_\perp$), quantum effects alone are found to single out \emph{one} most unstable mode where the classical cold case yields a continuum of them. The most unstable mode on Fig. \ref{fig:1}(c) is
\begin{equation}\label{eq:GRQuan}
   Z_{zM}\sim 1.06,~~Z_{xM}\sim 4.07,~~\mathrm{with}~~ \frac{\delta_M}{\omega_p} = 0.021
\end{equation}
which growth rate is close to the one of the cold classical system given by Eqs. (\ref{eq:GRcold},\ref{eq:GRColdNum}). Although quantum corrections are found to have a noticeable effect on the unstable spectrum, they do not alter the fastest growing mode.

\subsection{Beam temperature effects}
Beam temperature effects are very important, as evidenced on Fig. \ref{fig:1}(d).  In accordance with some recent conclusions drawn from a more elaborated kinetic relativistic model \cite{BretPRL2008}, the fastest growing collisionless mode is here oblique. Also, most of the unstable modes, filamentation included \cite{Molvig,Cary1981,Silva2002}, have been stabilized and the unstable spectrum is now restricted to a very narrow region of the $\mathbf{Z}$ plane. The maximum growth rate  has been reduced to 15\% of its cold classical value, and the fastest growing mode is now,
\begin{equation}\label{eq:GRQuanHot}
   Z_{zM}\sim 1.35,~~Z_{xM}\sim 1.08,~~\mathrm{with}~~ \frac{\delta_M}{\omega_p} = 0.0033.
\end{equation}
Note that the same calculation  canceling all quantum effects gives the very same result. So far, plasma degeneracy is found to affect poorly the dominant unstable mode.

\begin{figure}[t]
\begin{center}
 \includegraphics[width=\textwidth]{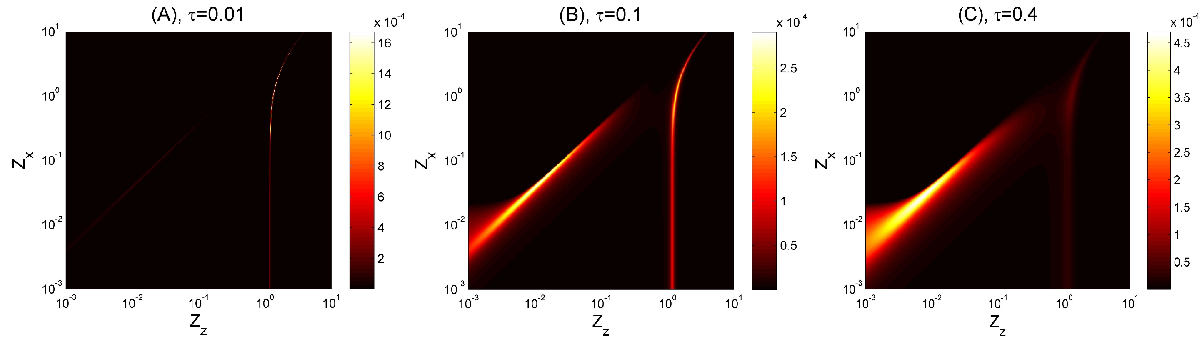}
\end{center}
\caption{(Color online) Growth rate of the whole unstable spectrum in terms of $\mathbf{Z}=\mathbf{k}v_b/\omega_p$ including all effects and for various collision frequencies.  Note the 3 orders of magnitude drop of the maximum growth with respect to Fig. \ref{fig:1}(a).}\label{fig:2}
\end{figure}

The growth rate map evolves dramatically  when collisions are accounted for. As observed on Fig. \ref{fig:2}(a-c), collisions strongly reduce the former largest growth rate. But the most remarkable evolution is the arising of oblique collisional unstable modes at much lower $k$. Indeed, the fastest growing mode and its growth rate is now,
\begin{equation}\label{eq:GRColl}
   Z_{zM}\sim 0.0078,~~Z_{xM}\sim 0.029,~~\mathrm{with}~~ \frac{\delta_M}{\omega_p} = 4.7\times 10^{-4}.
\end{equation}
Collisional effects are therefore essential in the present setting, and we now focus on them.

\section{\label{sec:Coll}Collisional effects}
The arising of collisional modes more unstable than the collisionless ones has already been mentioned in the literature for the filamentation instability \cite{Honda2004}. Still for the filamentation instability, the shift of the fastest growing mode to lower $k$'s has also been reported \cite{hao2008,cottrill2008,fiore}. Indeed, such a shift is desirable if the transition is to be made from the collisionless regime, where the filaments transverse size is of the order of the plasma skin depth, to the resistive regime \cite{gremillet2002,Honrubia2006}, where filaments transverse size is about the much larger \emph{beam} skin depth (at least in the diluted beam regime). What is evidenced here is the extension of this trend to the oblique modes which indeed govern the diluted relativistic regime \cite{BretPRL2008}.

We can monitor between Figs. \ref{fig:1}(d) and \ref{fig:2}(a-c) the ``birth'' of the collisional mode at much lower $k$, which eventually surpasses the collisionless fastest growing mode when $\tau$ is increased. On the one hand, collisions strongly limit the growth of the collisionless oblique modes while on the other hand, triggering some low $k$'s unstable modes.

We now focus on the transition dynamic. How does the system switches from one dominant mode to the other? Figs. \ref{fig:2}(a,b) display the unstable spectrum for intermediate collision frequencies between Fig. \ref{fig:1}(d), where $\tau=0$, and Fig. \ref{fig:2}(c), with $\tau=0.4$. Comparison suggests the transition happens here for $\tau < 0.1$. A finer evaluation yields $\tau=0.085$, which is much smaller than the 0.4 value considered here for FIS. Therefore, and according to the present model, the end of the beam path definitely pertains to the collisional regime, in the sense that unstable modes stemming from collisionality govern the system. Noteworthily, the transition is abrupt. We do not have one mode  evolving continuously from one regime to the other. Instead, as collisions are progressively ``switched-on'', dominant collisional unstable modes are progressively stabilized in one region of the unstable spectrum, while in a completely different region, collisional modes grow faster and faster. When some critical collision frequency is reached (here $\nu=0.085\omega_p$), the latters overcome the formers and the dominant mode ``jumps'' from the resonant part of the spectrum $Z_z\sim 1$ to a much lower $Z_z$ one. Figures \ref{fig:2}(b,c) show that the new modes are shifted down by some 2 orders of magnitude along the $Z_z$ and $Z_x$ axis. Remembering the beam to plasma density ratio is here $10^{-4}$, two orders of magnitude also separate the plasma skin depth from the beam one. The low $k$ modes found here generate patterns of the ``correct'' typical size, namely, the beam skin depth one.

\begin{table}

\caption{\label{tab:summary}Most unstable wave vector and its growth rate (in $\omega_p$ units) accounting for the various effects presented. Except in the first case, there is always one single most unstable mode.}
\begin{indented}
\item[]\begin{tabular}{@{}llll}
\br
Model                     &       $\delta_M$  &  $Z_{zM}$     &     $Z_{xM}$\\
\mr
Cold only                 &       0.022                  &   1        &   $\gg$ 5   \\
Bohm term                 &       0.022                  &   1        &   13        \\
All Quantum terms         &       0.021                  &   1.06     &   4.07    \\
Quantum + thermal         &       0.0033                 &   1.35     &   1.08    \\
Thermal only              &       0.0033                 &   1.35     &   1.08     \\
All effects + $\tau=0.4$  &       $4.7\times 10^{-4}$    &   0.0078   &   0.029   \\
\br
\end{tabular}
\end{indented}
\end{table}

Table \ref{tab:summary} allows for a global picture of the results gathered in Secs. \ref{sec:Q} \& \ref{sec:Coll}. Every effect added to the cold case calculation reduces the maximum growth rate. Quantum effects only slightly do so, while the beam thermal spread dramatically slows down the fastest growing mode. This most unstable mode remains on the very same branch on the dispersion equation as long as collisions are neglected. By setting $\tau\neq 0$, a new branch appears at much lower $k$, which overcomes the collisionless branch from $\tau = 0.085$. Furthermore, the hot and collisional system is about 3 orders of magnitude less unstable than its cold counterpart.

\section{Conclusion}
By implementing a relativistic fluid model for beam plasma interaction, we could assess the relative importance of collisions and quantum effects for the instabilities arising when a hot relativistic electron beam interacts with the dense core of a pre-ignited and pre-compressed FIS target. We consider the beam as relativistic, collisionless and classical. The plasma return current is degenerate, non-relativistic and collisional. Also, we computed the whole unstable spectrum in order to spot the fastest growing unstable mode, whether it is two-stream, filamentation or oblique-like. For the typical FIS parameters mentioned in Eqs. (\ref{eq:paramnum}), Table \ref{tab:summary} summarizes the growth rate of the fastest growing mode as well as its location accounting for the various effects.

Quantum effects result from the inclusion of the zero temperature Bohm pressure term, and of a quantum thermal correction. The Bohm term hardly modifies the unstable spectrum and the largest growth rate, as can be checked from  Table \ref{tab:summary} or  Figs. \ref{fig:1}(a,b). However, by damping large $k_\perp$ unstable modes, it singles out \emph{one} most unstable wave vector. This trend amplifies when including quantum thermal corrections. The normal component of the most unstable wave vector decreases even more, but the value of $\delta_M$ remains unchanged. Finally, calculations accounting for beam temperature and collisions are insensitive to the inclusion of any quantum effects. Although some kinetic treatment could be needed to settle the case, it seems that thermal and collision effects are definitely the dominant ones in FIS context whereas plasma degeneracy can be neglected. Furthermore, plasma temperature should rise as the beam starts depositing its energy, resulting in a even less degenerate plasma.

Including beam temperature brings about a major modification of the unstable spectrum (see Fig. \ref{fig:1}d). As expected from previous kinetic studies \cite{Silva2002,Cary1981}, filamentation modes are here completely damped. Even oblique modes close to filamentation have been stabilized and by virtue of the very small beam to plasma density ratio, the unstable spectrum has been reduced to a thin line starting from $Z_z\sim 1$. Both the location and the growth rate of the most unstable mode are affected, and the typical size of the generated patterns is the plasma skin depth.

Finally, and this may be the most important conclusion of this paper, the inclusion of a strong collision electron-ion frequency for the plasma return current triggers some very low $k$ oblique unstable modes. These longer wavelength modes overcome the collisionless ones as soon as $\nu > \nu_c=0.085\omega_p$. Note that such modes also exist along the perpendicular direction, but the dominant ones are oblique. Given the high collision frequency considered in our case ($0.4\omega_p$), the collisional regime should definitely be switched-on. The transition from one regime to the other is discontinuous when crossing the $\nu_c$ threshold because the two competing modes are located in remote places of the spectrum. A discontinuous evolution of the growth rate was already pointed out at $\nu=0$ by Molvig, in a kinetic analysis of the magnetized filamentation instability  \cite{Molvig}. In the present case, the growth rate is a continuous function of $\mathbf{k}$ and $\nu$, and the discontinuity arises only when considering the maximum growth rate over the $\mathbf{k}$ spectrum. Additionally, the discontinuity occurs here for $\nu=\nu_c\neq 0$.

By growing at much smaller $k$ (see Table \ref{tab:summary}), the collisional mode generate patterns which typical size is now the much larger \emph{beam} skin depth, reminiscent of the so-called resistive filamentation instability \cite{gremillet2002,Honrubia2006}. At the present stage, we think more studies are needed to identity the present collisional modes with this instability, but similarities are striking. A previous cold model \cite{BretPoPColl} accounting for the same collision frequency for both the beam and plasma electrons, failed to generate such modes. Furthermore, we checked that the present model  yields fast growing collisional \emph{filamentation} modes in the absence of beam temperature. This latter factor thus seems responsible for the obliqueness of the fastest growing mode, while the plasma  collisionality (together with the beam  \emph{non}-collisionality) seem to be the reasons for their very existence. It is probable that a certain degree of collision is acceptable in the beam before the collisional regime vanishes. Since the expression of the dielectric tensor is known, it should be possible to access the collisional modes analytically, gaining thus a deeper understanding of the transition uncovered here. Appropriate approximations are currently being developed allowing to simplify the dielectric tensor. Finally, beam temperature effects on the collisional modes need to be kinetically assessed to go beyond the present sub-relativistic temperature limit.

\section{Acknowledgements}
This work has been  achieved under projects FIS 2006-05389 of the
Spanish Ministerio de Educaci\'{o}n y Ciencia and PAI08-0182-3162 of
the Consejer\'{i}a de Educaci\'{o}n y Ciencia de la Junta de
Comunidades de Castilla-La Mancha.

\end{document}